\documentclass[twocolumn,english]{revtex4-2}
\usepackage[T1]{fontenc}
\usepackage[latin9]{inputenc}
\usepackage[a4paper]{geometry}
\usepackage{amsmath,amssymb,amsfonts}
\usepackage{comment}
\usepackage{kbordermatrix}
\renewcommand{\kbldelim}{(}
\renewcommand{\kbrdelim}{)}

\def\be{\begin{equation}}
\def\ee{\end{equation}}

\newcommand{\eq}[1]{\begin{equation}#1\end{equation}}
\newcommand{\eqs}[1]{\begin{equation}\begin{split}#1\end{split}\end{equation}}

\begin{document}
\widetext

\title{Self-dual gravity and color/kinematics duality in AdS$_4$}
\author{Arthur Lipstein $^{a}$}
\author{Silvia Nagy $^{a}$}
\affiliation{$^{a}$Department of Mathematical Sciences, Durham University, Durham, DH1 3LE, UK}

\begin{abstract}
We show that self-dual gravity in Euclidean four-dimensional Anti-de Sitter space (AdS$_4$) can be described by a scalar field with a cubic interaction written in terms of a deformed Poisson bracket, providing a remarkably simple generalisation of the Plebanski action for self-dual gravity in flat space. This implies a novel symmetry algebra in self-dual gravity, notably an AdS$_4$ version of the so-called kinematic algebra. We also obtain the 3-point interaction vertex of self-dual gravity in AdS$_4$ from that of self-dual Yang-Mills by replacing the structure constants of the Lie group with the structure constants of the new kinematic algebra, implying that self-dual gravity in AdS$_4$ can be derived from self-dual Yang-Mills in this background via a double copy. This provides a concrete starting point for defining the double copy for Einstein gravity in AdS$_4$ by expanding around the self-dual sector. Moreover, we show that the new kinematic Lie algebra can be lifted to a deformed version of the $w_{1+\infty}$ algebra, which plays a prominent role in celestial holography.
\end{abstract}

\maketitle

\section{Introduction}
Self-dual Yang-Mills (SDYM) and gravity (SDG) have provided a very fruitful setting for studying the mathematical structure of perturbative quantum gravity in asympotically flat background. For example, in lightcone gauge they can be described by very simple scalar theories \cite{Plebanski:1975wn,Bardeen:1995gk,Chalmers:1996rq,Prasad:1979zc,Dolan:1983bp,Parkes:1992rz,Cangemi:1996pf,Popov:1996uu,Popov:1998pc} which make various properties such as color/kinematics duality and the double copy manifest, as shown in \cite{Monteiro:2011pc} and further explored in \cite{Chacon:2020fmr,Campiglia:2021srh,Monteiro:2022nqt,Monteiro:2013rya,Cheung:2016prv,Chen:2019ywi,Elor:2020nqe,Armstrong-Williams:2022apo,Farnsworth:2021wvs,Skvortsov:2022unu,He:2015wgf,Berman:2018hwd,Nagy:2022xxs}. Color/kinematics duality is a relation between the color structures and kinematic numerators appearing in Feynman diagrams \cite{Bern:2008qj} which lies at the heart of the double copy relating gravity to the square of gauge theory, allowing one to reduce complicated calculations in the former to simpler calculations in the latter \cite{Kawai:1985xq,Bern:2010yg,Bern:2010ue}.

Another notable feature of SDYM and SDG is their integrability \cite{Ward:1977ta,Dunajski_1998,Prasad:1979zc,Dolan:1983bp,Popov:1996uu,Popov:1998pc,Park:1989vq,Husain:1993dp,Husain:1993dp}, which in the case of SDG may be linked to an infinite dimensional symmetry known as the $w_{1+\infty}$ algebra. This algebra is closely related to the kinematic algebra in SDG \cite{Monteiro:2022lwm} and may play a fundamental role in describing 4d quantum gravity in asympotically flat background via a two-dimensional conformal field theory (CFT) living on the sphere at null infinity, known as the celestial CFT \cite{Guevara:2021abz,Strominger:2021lvk,Ball:2021tmb,Adamo:2021lrv,Donnay:2022sdg,Pasterski:2020pdk,Casali:2020vuy,Donnay:2020guq,Casali:2020uvr,Mason:2022hly,Monteiro:2022xwq,Ball:2021tmb,Adamo:2019ipt,Strominger:2013jfa,He:2014laa,McLoughlin:2022ljp,Kampf:2023elx,Freidel:2021ytz}. Celestial CFT provides a framework to recast soft theorems of scattering amplitudes and their underlying asympotic symmetries in the language of 2d CFT.

The holographic description of quantum gravity is best understood in Anti-de Sitter space (AdS), where the dual description is provided by a CFT on the boundary \cite{Maldacena:1997re}. Furthermore, the study of boundary correlators in four dimensional Anti-de Sitter space (AdS$_4$) is relevant for cosmology after Wick rotating to four dimensional de Sitter space (dS$_4$) \cite{Strominger:2001gp,Maldacena:2002vr,McFadden:2009fg}. There has recently been a great deal of progress formulating color/kinematics duality and the double copy in (A)dS \cite{Armstrong:2020woi,Albayrak:2020fyp,Alday:2021odx,Diwakar:2021juk,Sivaramakrishnan:2021srm,Cheung:2022pdk,Herderschee:2022ntr,Drummond:2022dxd,Farrow:2018yni,Lipstein:2019mpu,Jain:2021qcl,Zhou:2021gnu,Armstrong:2022csc,4ptdouble} (and there are also the beginnings of a larger programme to extend the double copy to curved backgrounds \cite{Adamo:2017nia,Adamo:2018mpq,Bahjat-Abbas:2017htu,Carrillo-Gonzalez:2017iyj,Borsten:2019prq,Borsten:2021zir,Chawla:2022ogv}) although a systematic understanding is still lacking.

In this paper we set out to find a simple description of SDG in AdS$_4$ in order to gain a deeper understanding of how color/kinematics duality and the double copy work in this background. After generalising the self-duality equation to nonzero cosmological constant, we show that the solution for the metric can be elegantly written in terms of a scalar field obeying a simple generalisation of the equation of motion found long ago by Plebanski for SDG in flat space \cite{Plebanski:1975wn}. In particular it describes a scalar field with interactions encoded by a deformed Poisson bracket. From this we deduce that SDG can be derived from SDYM in this background by replacing the color algebra with a deformed kinematic algebra which reduces to the flat space one as the AdS radius goes to infinity. Even more surprisingly, we find that that this kinematic algebra can be lifted to a deformed version of the $w_{1+\infty}$ algebra suggesting exciting new connections between AdS/CFT and flat space holography. 

This paper is organised as follows. In section \ref{sdym}, we consider SDYM in AdS$_4$, which obeys the same equation of motion as flat space (the fact that we are working in AdS is only encoded by boundary conditions). In section \ref{sdg} we then look at SDG in AdS$_4$. We introduce an appropriate generalisation of the self-duality condition, and use it to extract a simple Plebanski-like scalar equation. This exhibits a modified Poisson bracket and double copy structure. In section \ref{ckd} we show that SDG in this background encodes a new kinematic algebra and can be obtained by combining this with the flat space kinematic algebra via an asymmetric double copy. In section \ref{winf}, we then lift the new kinematic algebra to a deformed $w_{1+\infty}$ algebra. We present our conclusions in \ref{conclusion}. There is also an appendix providing more details of the derivation of our SDG solution.

\section{Self-dual Yang-Mills} \label{sdym}

We will consider four dimensional Euclidean AdS$_4$ with unit radius in the Poincar\'e patch:
\eq{
ds_{\rm{AdS}}^{2}=\frac{dt^{2}+dx^{2}+dy^{2}+dz^{2}}{z^{2}},
\label{metric}
}
where $0<z<\infty$ is the radial coordinate. In a general background, the self-duality constraint for Yang-Mills theory (YM) reads
\eq{
F_{\mu\nu}=\frac{\sqrt{g}}{2}\epsilon_{\mu\nu\rho\lambda}F^{\rho\lambda},
}
where $g_{\mu \nu}$ is the background metric and $g$ is its determinant. In conformally flat spaces, such as AdS$_4$, this just reduces to the self-duality constraint in $\mathbb{R}^4$, since four dimensional YM is classically scale invariant. Indeed, for the metric in \eqref{metric},  the $\sqrt{g}$ yields a factor of $z^{-4}$ while the inverse metrics used to raise indices of the field strength give $z^{4}$, so these factors just cancel out. 

We work in the so-called light-cone coordinates:
\eqs{
u=it+z,\quad v=it-z,\\
w=x+iy,\quad \bar{w}=x-iy,
\label{lccoords}
}
in which the metric is given by
\be 
ds_{\rm{AdS}}^{2}=\frac{4\left(dw\,d\bar{w}-du\,dv\right)}{\left(u-v\right)^{2}},
\ee
and $\epsilon_{uvw\bar{w}}=-1$. The non-trivial self-duality contraints can then be written as
\eqs{
F_{uw}=F_{v\bar{w}}=0,\quad F_{uv}=F_{w\bar{w}}.
}
Following \cite{Bardeen:1995gk}, we will also impose lightcone gauge $A_{u}=0$. We then find that the self-duality constraints are solved by \cite{explEq6}
\eq{\label{YM flat in comp}
A_{w}=0,\quad A_{\bar{w}}=\partial_{u}\Phi, \quad A_{v}=\partial_{w}\Phi,
}
where $\Phi$ is a scalar field in the adjoint representation which satisfies the following equation of motion:
\eqs{
\Box_{\mathbb{R}^{4}}\Phi+i\left[\partial_{u}\Phi,\partial_{w}\Phi\right]=0,
}
where $\Box_{\mathbb{R}^{4}}=-\partial_{u}\partial_{v}+\partial_{w}\partial_{\bar{w}}$. 
This can in turn be derived from the following Lagrangian by introducing a Lagrange multiplier field $\bar{\Phi}$:
\eqs{
\mathcal{L}_{\rm{SDYM}}={\rm Tr}\left[\bar{\Phi}\left(\Box_{\mathbb{R}^{4}}\Phi+i\left[\partial_{u}\Phi,\partial_{w}\Phi\right]\right)\right].
\label{ymlag}
}
This is the same action that was previously derived for SDYM in flat space \cite{Bardeen:1995gk,Chalmers:1996rq,Prasad:1979zc,Dolan:1983bp,Parkes:1992rz,Cangemi:1996pf,Popov:1998pc} since AdS$_4$ can be conformally mapped to half of $\mathbb{R}^{4}$. On the other hand, since there is a boundary at $z=0$, momentum along the $z$ direction will not be conserved, which will become visible when computing boundary correlators in this background. We save a detailed analysis for future work.

With a view to the gravity formulation, we find it useful to split the spacetime coordinates as
\be
x^i=(u,w),\qquad y^\alpha=(v,\bar{w}),
\ee
and introduce the operators
\be\label{pidef}
\Pi _\alpha=(\Pi_v,\Pi_{\bar{w}})=(\partial_w,\partial_u),
\ee
which allows us to write the gauge field in \eqref{YM flat in comp} as
\be
A_i=0,\quad A_\alpha=\Pi_\alpha \Phi .
\ee 
Finally, we define the Poisson bracket \cite{Monteiro:2011pc}
\be\label{poisson_def}
 \{f,g\}:= \partial_{w}f\partial_{u}g-\partial_{u}f\partial_{w}g=\varepsilon^{\alpha\beta} \Pi_\alpha f \Pi_\beta g,
\ee
and notice that it appears naturally in the scalar equation of motion:
\be
\Box_{\mathbb{R}^{4}}\Phi-\frac{i}{2}[\{\Phi,\Phi\}]=0,
\ee
where we introduced the notation
\be
[\{f,g\}]= \varepsilon^{\alpha\beta} \left[\Pi_\alpha f,\Pi_\beta g\right].
\ee
Since the Poisson bracket obeys the Jacobi identity, it is the kinematic analogue of a commutator encoding the color algebra. Hence, SDYM manifestly exhibits color-kinematics duality since it posesses both a commutator and a Poisson bracket structure. The double copy involves replacing color structures with kinematic structures, mapping gauge theoretic quantities into gravitational ones. As we will see in the next section, replacing the commutator with another Poisson bracket yields a scalar action for self-dual gravity.

\section{Self-dual Gravity} \label{sdg}

In this section, we will first review SDG in flat background, as first derived in \cite{Plebanski:1975wn} and then describe the generalisation to AdS$_4$.

\subsection{Self-duality in asymptotically flat gravity}
In asymptotically flat gravity, the self-duality condition is given by 
\be \label{sd_cond_flat}
R_{\mu \nu \rho\sigma} = \tfrac{1}{2} \sqrt{g} \epsilon_{\mu \nu}^{\phantom{\mu \nu }\eta \lambda} R_{\eta \lambda\rho\sigma}.
\ee 
The above is the appropriate form of the condition in Euclidean signature. One can go to Lorenzian signature by rescaling the right-hand-side by a factor of $i$, and the coordinates apropriately. Crucially, the self-duality condition encodes both the equations of motion and the algebraic Bianchi identity for the Riemann tensor, which can be seen by contracting two of the indices in \eqref{sd_cond_flat} to get
\be
R_{\mu\rho}= \tfrac{1}{2}\epsilon_\mu^{\phantom{\mu}\sigma\eta\lambda}R_{\eta \lambda\rho\sigma}=0.
\ee
Writing the metric as 
\be \label{defg}
ds^2=dw\,d\bar{w}-du\,dv + h_{\mu\nu}\,dx^\mu dx^\nu,
\ee
we find that \eqref{sd_cond_flat}, together with the light-cone gauge choice $h_{u\mu}=0$ leads to 
\be 
h_{i\mu}=0,\quad h_{\alpha \beta} = \Pi_\alpha \Pi_\beta\phi \label{hphi},
\ee
with $\Pi_\alpha$ as defined in the YM sector \eqref{pidef} and the scalar $\phi$ satisfying  
\be \label{eomphi}
\Box_{\mathbb{R}^{4}} \phi - \{\{\phi,\phi\}\}=0,
\ee
where we introduced the notation
\be 
\{\{f,g\}\}=\frac{1}{2}\varepsilon^{\alpha\beta} \{\Pi_\alpha f, \Pi_\beta g  \},
\ee
and $\{,\}$ is the Poisson bracket introduced previously in \eqref{poisson_def}. This then alows us to give elegant double copy rules in the self-dual sector via \cite{Monteiro:2011pc}:
\be
 \Phi \to \phi, \qquad \frac{i}{2}[\{ \ , \ \}] \to  \{\{ \ , \ \}\}.
\ee

\subsection{Self-duality in AdS$_4$ gravity}
We wish to generalise the self-duality condition to AdS$_4$. To this end, we introduce the tensor:
\be \label{def_T_tensor}
T_{\mu\nu\rho\sigma} = R_{\mu\nu\rho\sigma} -\tfrac{1}{3}\Lambda (g_{\mu\rho}g_{\nu\sigma}-g_{\nu\rho}g_{\mu\sigma}),
\ee
where $\Lambda$ is the cosmological constant. We now define our duality relation as
\be 
\label{sd_cond_AdS}
T_{\mu \nu \rho\sigma} = \tfrac{1}{2}\sqrt{g}\epsilon_{\mu \nu}^{\phantom{\mu \nu }\eta \lambda} T_{\eta \lambda\rho\sigma}.
\ee
Upon contracting with $g^{\nu\sigma}$ we find
\be
R_{\mu\rho}-\Lambda g_{\mu\rho}=\tfrac{1}{2}\sqrt{g}\epsilon_\mu^{\phantom{\mu}\sigma\eta\lambda}R_{\eta \lambda\rho\sigma}=0,
\ee
where the left-hand-side gives the Einstein equation with a cosmological constant in the absence of matter sources
\be \label{lambda_equations}
R_{\mu\nu}=\Lambda g_{\mu\nu},\quad R=4\Lambda \, ,
\ee 
and the right-hand-side is again the algebraic Bianchi identity for the Riemann tensor.

As a side comment, we note another way in which $T_{\mu \nu \rho\sigma}$ is the natural generalisation of the Riemann tensor appearing in \eqref{sd_cond_flat} to spaces with a non-zero cosmological constant. To this end, it helps to consider the Weyl tensor in 4 dimensions:
\be 
C_{\mu\nu}^{\phantom{\mu\nu}\rho\sigma}=R_{\mu\nu}^{\phantom{\mu\nu}\rho\sigma}
-2R_{[\mu}^{\phantom{\mu}[\rho}g_{\nu]}^{\phantom{\nu}\sigma]}
+\frac{1}{3}R g_{[\mu}^{\phantom{\mu}[\rho}g_{\nu]}^{\phantom{\nu}\sigma]}
\ee
In asymptotically flat spaces, upon application of the vacuum equation of motion $R_{\mu\nu}=R=0$, we get the well-known result that the Weyl tensor becomes equal to the Riemann tensor. However, in the presence of a cosmological constant, the relevant equations are those in \eqref{lambda_equations}. Upon plugging these into the Weyl tensor, we recover exactly the form of $T_{\mu \nu \rho\sigma}$ from \eqref{def_T_tensor}\cite{aknMP}. This result is also natural from the spinorial formulation of tensors in General Relativity, where the so-called Weyl spinors, arising from the Weyl tensor, encode the self-dual and anti-self-dual degrees of freedom, upon applying the equations of motion. 

We will now specialise to a background with cosmological constant $\Lambda=-3$, corresponding to AdS$_4$ background with unit radius.

\subsection{Solution}

In this section, we will show that the solution to the self-duality constraint in AdS$_4$ is a remarkably simple generalisation of the flat space one in \eqref{eomphi} when written in terms of a deformed Poisson bracket. Let us begin by introducing the modified Poisson bracket:
\be
\begin{aligned}
\left\{ f,g\right\} _{*}=\left\{ f,g\right\} +\tfrac{2}{u-v}\left(f\partial_{w}g-g\partial_{w}f\right).
\label{poissone}
\end{aligned}
\ee
Using a deformation of the operators \eqref{pidef}
\be
\tilde{\Pi} = (\tilde{\Pi}_v,\tilde{\Pi}_{\bar{w}})=\left(\partial_w,\partial_u-\tfrac{4}{u-v}\right),
\ee
we can write \eqref{poissone} as 
\be \label{def_mod_poiss}
\left\{ f,g\right\} _{*}=\frac{1}{2}\varepsilon^{\alpha\beta}(\Pi_\alpha f\tilde{\Pi}_\beta g-\Pi_\alpha g\tilde{\Pi}_\beta f).
\ee
In this form, the Poisson bracket which previously appeared in flat space can be recovered simply by replacing $\tilde{\Pi}$ with its undeformed version $\Pi$: 
\eq{
\left\{  f,g \right\} =\left.\left\{ f,g \right\} _{*}\right|_{\tilde{\Pi}\rightarrow\Pi}.
} 
We also observe the following relation between the brackets:
\be
\left\{  f,g \right\}_*=(u-v)^4 \left\{ \frac{f}{(u-v)^2},\frac{g}{(u-v)^2}\right\}. 
\ee

Let us proceed to solve the self-duality equation in \eqref{sd_cond_AdS}. First we make the following general ansatz: 
\eqs{
ds^{2}=\frac{4\left(dw\,d\bar{w}-du\,dv+h_{\mu\nu}\,dx^{\mu}dx^{\nu}\right)}{\left(u-v\right)^{2}},
\label{ansatz}
}
where $h_{\mu \nu}$ are unfixed functions. Imposing lightcone gauge $h_{u \mu}=0$, we then find the following simple solution:
\eq{
h_{i\mu}=0,\quad h_{\alpha\beta}=\Pi_{(\alpha}\tilde{\Pi}_{\beta)}\phi,
\label{hsol}
}
where $\phi$ is a scalar field satisfying the following equation of motion:
\eq{
\frac{1}{u-v}\Box_{\mathbb{R}^{4}}\left(\frac{\phi}{u-v}\right)-\left\{ \left\{ \frac{\phi}{u-v},\frac{\phi}{u-v}\right\} \right\} _{*}=0,
\label{sdgeom}
}
where the modified double Poisson is defined as follows:
\be 
\left\{ \left\{ f,g\right\} \right\} _{*} =\frac{1}{2}\varepsilon^{\alpha\beta}\{\Pi_\alpha f, \Pi_\beta g  \}_{*},
\ee
with $\{,\}_{*}$ defined in \eqref{def_mod_poiss}. Setting $f=g$, this becomes
\eqs{
\left\{ \left\{ f,f\right\} \right\} _{*} &= \partial_{w}^{2}f\partial_{u}^{2}f-\left(\partial_{u}\partial_{w}f\right)^{2} \\  &\qquad+\frac{2}{u-v}\left(\partial_{w}f\partial_{u}\partial_{w}f-\partial_{u}f\partial_{w}^{2}f\right).
}
We provide details of how to solve the self-duality equations in Appendix \ref{details}. 

Hence, the equation of motion in \eqref{sdgeom} provides a natural generalisation of the equation of motion for SDG in flat space given in \eqref{eomphi}. In particular, it exhibits an asymetric double copy structure 
\be 
 \Phi \to \frac{\phi}{u-v}, \qquad \frac{i}{2}[\{ \ , \ \}] \to  \{\{ \ , \ \}\}_{*} \ ,
\ee
up to a rescaling of the kinetic term, which will be explored further in the next section. After some algebra, the equation of motion in \eqref{sdgeom} can also be written as follows: 
\eqs{\label{eom with mass ADS}
\sqrt{g}\left(-\Box_{\rm{AdS}}+m^{2}\right)\phi+4\left\{ \left\{ \frac{\phi}{u-v},\frac{\phi}{u-v}\right\} \right\} _{*}=0,
}
where $\Box_{\rm{AdS}}\phi=g^{-1/2}\partial_{\mu}\left(\sqrt{g} g^{\mu\nu}\partial_{\nu}\phi\right)$ with $g_{\mu\nu}$ the background AdS$_4$ metric, and $m^{2}=-2$ corresponding to a conformally coupled scalar in AdS$_4$. Recall that a conformally coupled scalar field in AdS$_{d+1}$ has mass $m^{2}=-\frac{d^{2}-1}{4}$ and can be mapped to a massless scalar field in flat space by a Weyl transformation \cite{Sonego:1993fw}. The equation of motion in \eqref{eom with mass ADS} is in turn encoded by the following Larangian: 
\eq{
\mathcal{L}_{\rm{SDG}}=\sqrt{g}\bar{\phi}\left(-\Box_{{\rm {AdS}}}+m^{2}\right)\phi+4\bar{\phi}\left\{ \left\{ \frac{\phi}{u-v},\frac{\phi}{u-v}\right\} \right\} _{*},
\label{grlag}
}
where $\bar{\phi}$ is a Lagrange multiplier field. 

Finally, \eqref{sdgeom} admits the following solutions, which are related to planewave solutions by a Weyl rescaling:
\eq{
\phi=(u-v) e^{i k\cdot x},
\label{solutions}
}
where $k\cdot x \equiv u k_u+ v k_v+w k_w+\bar{w} k_{\bar{w}}$  is the flat space inner product and $k_{u}k_{v}-k_{w}k_{\bar{w}}=0$ (we refer to this as the on-shell condition). Note that the momenta are complex since we are working in Euclidean signature. Since  there is a boundary at $z=0$, momentum along the $z$ direction will not be conserved and the natural observables are boundary correlators Fourier transformed to momentum space \cite{Raju:2011mp,Maldacena:2011nz,Bzowski:2013sza,Bzowski:2015pba}.

\section{Color/kinematics duality} \label{ckd}

It is straightforward to read off Feynman rules from the Lagrangians in \eqref{ymlag} and \eqref{grlag}. First we expand the scalar fields in the SDYM action as $\Phi=\Phi^a T^a$, where $T^a$ generators of the gauge group satisfying ${\rm Tr}\left(T^{a}T^{b}\right)=\delta^{ab}$ and $\left[T^{a},T^{b}\right]=i f^{abc}T^{c}$. Using on-shell plane wave external states for SDYM and external states of the form \eqref{solutions} for SDG, we obtain the following 3-point vertices (which would be relevant when computing three-point boundary correlators): 
\eqs{
V_{\rm{SDYM}} &=\frac{1}{2}X\left(k_{1},k_{2}\right)f^{a_{1}a_{2}a_{3}},\\V_{\rm{SDG}}&=\frac{1}{2}X\left(k_{1},k_{2}\right)\tilde{X}\left(k_{1},k_{2}\right),
}
where 
\eqs{
X\left(k_{1},k_{2}\right)&=k_{1u}k_{2w}-k_{1w}k_{2u},\\ \tilde{X}\left(k_{1},k_{2}\right)&=X\left(k_{1},k_{2}\right)-\frac{2i}{u-v}\left(k_{1}-k_{2}\right)_{w}.
}

The objects $X$ and $\tilde{X}$ obey Jacobi identities analogous to $f^{a_{1}a_{2}a_{3}}$ and can therefore be thought of as structure constants of kinematic Lie algebras:
\eqs{
0&=X\left(k_{1},k_{2}\right)X\left(k_{3},k_{1}+k_{2}\right)+{\rm cyclic}\\
&=\tilde{X}\left(k_{1},k_{2}\right)\tilde{X}\left(k_{3},k_{1}+k_{2}\right)+{\rm cyclic}.
\label{jacobikin}
}
These relations do not rely on momentum conservation, and encode color/kinematics duality. Moreover, we find that the SDG vertex can be obtained from the SDYM one by replacing the color structure constant with the deformed kinematic structure contant:
\eq{
f^{a_{1}a_{2}a_{3}}\rightarrow\tilde{X}\left(k_{1},k_{2}\right),
}
which encodes the double copy. Whereas in flat background there is only one kinematic algebra and the SDG vertex is obtained by simply squaring $X$ \cite{Monteiro:2011pc}, in AdS$_4$ there are two distinct kinematic algebras and SDG arises from an asymmetrical double copy. 

The kinematic structure constants naturally arise from Poisson brackets on plane waves:
\eqs{
\left\{ e^{ik_{1}\cdot x},e^{ik_{2}\cdot x}\right\} &=X\left(k_{1},k_{2}\right)e^{i\left(k_{1}+k_{2}\right)\cdot x},\\
\left\{ e^{ik_{1}\cdot x},e^{ik_{2}\cdot x}\right\} _{*} &=\tilde{X}\left(k_{1},k_{2}\right)e^{i\left(k_{1}+k_{2}\right)\cdot x},
}
where $k \cdot x$ is defined below \eqref{solutions}. Note that when we plug the solutions in \eqref{solutions} into the deformed Poisson bracket in \eqref{grlag}, this is indeed equivalent to acting on plane waves since we divide by the conformal factor $(u-v)$. The kinematic Jacobi identity in \eqref{jacobikin} is a consequence of the following general property of the deformed Poisson bracket:
\be 
\left\{f,\left\{ g, h\right\} _{*}\right\} _{*}+\left\{g,\left\{ h, f\right\} _{*}\right\} _{*}+\left\{h,\left\{ f, g\right\} _{*}\right\} _{*}=0.
\ee
for arbitrary functions $f,g$ and $h$. Note that the deformed Poisson bracket satisfies a deformed Leibniz rule:
\be 
\left\{ \tfrac{fg}{(u-v)^2},h\right\} _{*}=\tfrac{1}{(u-v)^2}f\left\{ g,h\right\} _{*}+\tfrac{1}{(u-v)^2}g\left\{ f,h\right\} _{*},
\ee
or alternatively
\be
\left\{ fg,h\right\} _{*}=f\left\{ g,h\right\} _{*}+g\left\{ f,h\right\} _{*}-\frac{2fg\partial_{w}h}{u-v},
\ee
although this does not play an important role in our analysis.

\section{$w_{1+\infty}$ algebras} \label{winf}

As shown in \cite{Monteiro:2022lwm}, the kinematic algebra which appears in SDG can be lifted to a $w_{1+\infty}$ algebra, which plays an important role in the study of scattering amplitudes in the context of celestial CFT \cite{Strominger:2021lvk}. In particular, the $w_{1+\infty}$ algebra contains the extended BMS algebra underlying soft graviton theorems of scattering amplitudes \cite{Strominger:2013jfa,He:2014laa}. In this section, we will follow similar steps to those in \cite{Monteiro:2022lwm} to show that the deformed kinematic algebra derived in the previous section can be lifted to a deformed $w_{1+\infty}$ algebra.

For an on-shell state, the momentum satisfies $k_{\bar{w}}/k_{u}=k_{v}/k_{w}=\rho$, where $\rho$ is some number. It is then possible to expand an on-shell plane wave as follows:
\eq{
e^{ik\cdot x}=\sum_{a,b=0}^{\infty}\frac{\left(ik_{u}\right)^{a}\left(ik_{w}\right)^{b}}{a!b!}\mathfrak{e}_{ab},
}
where $\mathfrak{e}_{ab}=\left(u+\rho \bar{w}\right)^{a}\left(w+ \rho v\right)^{b}$. This is naturally interpreted as an expansion in soft momenta. Letting $w_{m}^{p}=\frac{1}{2}\mathfrak{e}_{p-1+m,p-1-m}$ and plugging this into the Poisson brackets in \eqref{poisson_def} and \eqref{poissone} then gives
\eqs{
\left\{ w_{m}^{p},w_{n}^{q}\right\} &=\left(n(p-1)-m(q-1)\right)w_{m+n}^{p+q-2},\\
\left\{ w_{m}^{p},w_{n}^{q}\right\} _{*}&=\left\{ w_{m}^{p},w_{n}^{q}\right\} +\frac{(m+q-p-n)}{u-v}w_{m+n+1/2}^{p+q-3/2}.
}
We recognize the first line as the $w_{1+\infty}$ algebra \cite{Fairlie:1990wv,Strominger:2021lvk}, and the second line appears to be a deformed version of this algebra. We recognize the first line as the $w_{1+\infty}$ algebra \cite{Fairlie:1990wv,Strominger:2021lvk}, and the second line appears to be a deformed version of this algebra.  In the limit where $z=(u-v) \rightarrow \infty$ (which corresponds to the flat space limit), the deformation vanishes. This suggests that self-dual gravity in AdS$_4$ is integrable.

Constant deformations of the $w_{1+\infty}$ algebra have been classified in \cite{Pope:1991ig,Bittleston:2023bzp,Bu:2022iak,Etnigof:2020xx}, however we note that our deformation falls outside of this classification, since it depends on $u-v$.

\section{Conclusion} \label{conclusion}

We have shown that SDG in AdS$_4$ can be described by a scalar field whose interactions are encoded by a deformed Poisson bracket, providing a surprisingly simple generalisation of the Plebanski action for SDG in flat space. Our action implies a new kinematic algebra dual to the color algebra appearing in SDYM, which is a deformation of the flat space kinematic algebra. Moreover, the new kinematic algebra can be lifted to a deformation of the $w_{1+\infty}$ algebra, implying a new relation between AdS/CFT and flat space holography which extends beyond the flat space limit. Indeed, to our knowledge $w_{1+\infty}$ symmetry was not previously identified in the context of AdS/CFT. It would be interesting to see how our SDG action compares to previous proposals in \cite{Ward:1980am,Przanowski:1983xpa,Krasnov:2016emc,Krasnov:2021cva,Krasnov:2017dww,Krasnov:2021nsq,Krasnov:2022mvn,Herfray:2022prf,Neiman:2023bkq}, as well as how it can be realised in twistor space \cite{Adamo:2021bej,Adamo:2021lrv,Mason:2022hly}. In particular, scalar equations were proposed long ago in \cite{Przanowski:1983xpa} and more recently in \cite{Neiman:2023bkq}, although they appear to be nontrivially related to ours and the deformed Poisson structure is not manifest in those formulations. Moreover, it would be interesting to generalise our approach to other conformally flat backgrounds. 

There are a number of other directions for future study. Perhaps the most immediate task is to compute tree-level boundary correlators of SDYM and SDG in AdS$_4$ and investigate how they encode color/kinematics duality and $w_{1+\infty}$ symmetry. In doing so, we must take into account the fact that momentum along the radial direction is not conserved and that the bulk-to-bulk propagators must satisfy nontrivial boundary conditions as a result of the boundary at $z=0$. Note that the classical solutions in \eqref{solutions} correspond to bulk-to-boundary propagators and can be mapped to plane waves via a Weyl transformation. One slightly nonstandard aspect of these calculations will be the need to work in lightcone gauge, since previous treatments usually worked in axial gauge \cite{Raju:2011mp,Maldacena:2011nz}. We can then investigate if the correlators exhibit universal behaviour in the soft or collinear limit, analogous to those in flat space, and explore how this is encoded by the $w_{1+\infty}$ symmetry. Recent work relating soft theorems to Ward identities in 3d CFT may be of use in this regard \cite{Banerjee:2022oll,Donnay:2022aba,deGioia:2023cbd,Bagchi:2023fbj,Saha:2023hsl}. In flat space, the scattering amplitudes of SDYM and SDG are 1-loop exact rational functions \cite{Bern:1993qk,Bern:1998xc} which also exhibit color/kinematics duality \cite{Boels:2013bi}. It would be very interesting to see if any of these properties extends to loop-level boundary correlators in AdS$_4$.

As mentioned above, we can obtain SDG in AdS$_4$ from an asymmetrical double copy by combining the flat space kinematic algebra (which appears in SDYM) with a deformed kinematic algebra. It would be interesting to see what gravitational theory arises from squaring the deformed kinematic algebra, or alternatively what gauge theory arises from combining the deformed kinematic algebra with the color algebra. Our approach may also provide a framework for defining color/kinematics duality and the double copy in Einstein gravity via an expansion around the self-dual sector. Indeed, the 4-point tree-level wavefunction coefficient for gravitons in dS$_4$ (which can be obtained by analytic continuation from AdS$_4$) can be deduced from an ansatz resembling an asymmetric double copy with deformed kinematic numerators \cite{4ptdouble}. 

Finally, and perhaps most ambitiously, it would be interesing to identify the CFT dual to SDG in AdS$_4$. Given that the bulk theory may have an infinite dimensional symmetry it seems very likely that it is integrable, and it should be possible to prove this by generalising the arguments in \cite{Prasad:1979zc,Dolan:1983bp,Popov:1996uu,Popov:1998pc,Park:1989vq,Husain:1993dp,Husain:1993dp,Campiglia:2021srh}. SDG in AdS$_4$ may therefore provide an exactly solvable toy model of AdS/CFT. Moreover, introducing a Moyal deformation analogous to the one recently implemented for SDG in flat space \cite{Monteiro:2022xwq} may describe a chiral higher spin theory in AdS$_4$ \cite{Sharapov:2022awp}. There are many exciting avenues which we hope to explore in the future. 

\begin{acknowledgments}
We thank Roberto Bonezzi, Simon Heuveline, Gregory Korchemsky, Kirill Krasnov, Lionel Mason, Ricardo Monteiro, Malcolm Perry, Andrea Puhm and David Skinner for useful discussions. AL is supported by the Royal Society via a University Research Fellowship. SN is supported in part by STFC consolidated grant T000708.
\end{acknowledgments}



\appendix

\section{Details of the gravity solution} \label{details}

In this Appendix, we will explain how to derive \eqref{sdgeom}.  Setting $h_{u \mu}=0$ in \eqref{ansatz}, we find that the determinant of the metric depends non-trivially on all components of $h_{w \mu}$ except for $h_{w v}$. Since the square root of the determinant appears on the right-hand-side of \eqref{sd_cond_AdS} but there is no square root on the left-hand-side, we set these components to zero. The self-duality equations then imply that $h_{w v}=0$. We may therefore consider the following ansatz:
\eqs{
ds^{2}=\frac{4\left(dw\,d\bar{w}-du\,dv+h_{\alpha\beta}\,dy^\alpha dy^\beta\right)}{\left(u-v\right)^{2}}.
\label{ansatz2}
}
Using this ansatz, we may write the self-duality constraint in \eqref{sd_cond_AdS} as
\eq{
\Delta_{\mu\nu\rho\sigma}=-T_{\mu\nu\rho\sigma}+\frac{2}{(u-v)^{2}}\epsilon_{\mu\nu\eta\lambda}T_{\,\,\,\,\,\,\,\,\rho\sigma}^{\eta\lambda},
}
where the left-hand-side must vanish. We then make an ansatz for $h_{\alpha \beta}$ in terms of derivatives of a scalar field and inverse powers of $(u-v)$ which reduces to \eqref{hphi} as $(u-v)\rightarrow \infty$. We then find that the following choice of coefficients in this ansatz causes most of the components of $\Delta$ to trivially vanish:
\eqs{
h_{vv} &=\partial_{w}^{2}\phi, \quad h_{v\bar{w}} =\partial_{u}\partial_{w}\phi-\frac{2}{u-v}\partial_{w}\phi, \\
h_{\bar{w}\bar{w}} &=\partial_{u}^{2}\phi-\frac{4}{u-v}\partial_{u}\phi+\frac{4}{(u-v)^{2}}\phi.
\label{hsol2}
}
This is indeed equivalent to \eqref{hsol}.

Furthermore, we find that the nonzero components of $\Delta$ can be written as follows:
\eqs{
\Delta_{vuwv}&=-\Delta_{\bar{w}w\bar{w}v}=\frac{4}{(u-v)^{2}}\partial_{w}{\rm eom},\\
\Delta_{\bar{w}vwv}&=-\frac{4}{(u-v)^{2}}\partial_{w}^{2}{\rm eom},\\
\Delta_{\bar{w}vvu}&=\frac{4}{(u-v)^{3}}\left((u-v)\partial_{u}\partial_{w}-\partial_{w}\right){\rm eom},\\
\Delta_{\bar{w}wwu}&=\frac{4}{(u-v)^{4}}\left((u-v)^{2}\partial_{u}^{2}-4(u-v)\partial_{u}+6\right){\rm eom},\\
\Delta_{\bar{w}vw\bar{w}}&=\frac{4}{(u-v)^{3}}\left((u-v)\partial_{u}\partial_{w}-3\partial_{w}\right){\rm eom},\\
\Delta_{\bar{w}v\bar{w}v}&=-\frac{4}{(u-v)^{4}}\left[(u-v)^{2}\left(\partial_{w}\partial_{\bar{w}}-\partial_{u}\partial_{v}\right)\right.\\
&\qquad \left.+(u-v)\left(\partial_{u}+3\partial_{v}\right)-2\partial_{w}\phi\partial_{w}+2\right]{\rm eom},
}
where
\eqs{
{\rm eom}&=\left(\partial_{u}\partial_{v}-\partial_{w}\partial_{\bar{w}}\right)\phi+\frac{1}{u-v}\left(\partial_{u}-\partial_{v}\right)\phi-\frac{2\phi}{(u-v)^{2}}\\
&\qquad+\left(h_{vv}h_{\bar{w}\bar{w}}-h_{v\bar{w}}^{2}+\frac{1}{(u-v)^{2}}\left(\partial_{w}\phi\right)^{2}\right).
\label{eom2}
}
Hence, all the components of $\Delta$ will vanish if we impose $\rm{eom}=0$, which can be interpreted as the equation of motion for the scalar field.

Note that the quantity $\rm{eom}$ can be written as
\eq{
{\rm eom={\rm kin}+{\rm pot}},
}
where
\eqs{
{\rm kin}&=\frac{\left(u-v\right)^{2}}{4}\sqrt{g}\left(-\Box_{{\rm {AdS}}}+m^{2}\right)\phi,\\
{\rm pot}&=\left(u-v\right)^{2}\text{\ensuremath{\left\{  \left\{  \frac{\phi}{u-v},\frac{\phi}{u-v}\right\}  \right\}  _{*}}},
}
where $\Box_{\rm{AdS}}\phi=g^{-1/2}\partial_{\mu}\left(\sqrt{g} g^{\mu\nu}\partial_{\nu}\phi\right)$ with $g_{\mu\nu}$ the background AdS$_4$ metric, and $m^{2}=-2$ (corresponding to a conformally coupled scalar in AdS$_4$). The kinetic and potential terms correspond to the first and second lines of \eqref{eom2}, respectively. This gives the equation of motion \eqref{eom with mass ADS}, which is equivalent to \eqref{sdgeom}.




\begin{thebibliography}{99}

\bibitem{Plebanski:1975wn}
J.~F.~Plebanski,
``Some solutions of complex Einstein equations,''
J. Math. Phys. \textbf{16} (1975), 2395-2402
doi:10.1063/1.522505

\bibitem{Bardeen:1995gk}
W.~A.~Bardeen,
``Selfdual Yang-Mills theory, integrability and multiparton amplitudes,''
Prog. Theor. Phys. Suppl. \textbf{123} (1996), 1-8
doi:10.1143/PTPS.123.1

\bibitem{Chalmers:1996rq}
G.~Chalmers and W.~Siegel,
``The Selfdual sector of QCD amplitudes,''
Phys. Rev. D \textbf{54} (1996), 7628-7633
doi:10.1103/PhysRevD.54.7628
[arXiv:hep-th/9606061 [hep-th]].

\bibitem{Prasad:1979zc}
M.~K.~Prasad, A.~Sinha and L.~L.~Wang,
``Nonlocal Continuity Equations for Selfdual SU($N$) {Yang-Mills} Fields,''
Phys. Lett. B \textbf{87} (1979), 237-238
doi:10.1016/0370-2693(79)90972-9

\bibitem{Dolan:1983bp}
L.~Dolan,
``Kac-moody Algebras and Exact Solvability in Hadronic Physics,''
Phys. Rept. \textbf{109} (1984), 1
doi:10.1016/0370-1573(84)90134-0

\bibitem{Popov:1996uu}
A.~D.~Popov, M.~Bordemann and H.~Romer,
``Symmetries, currents and conservation laws of selfdual gravity,''
Phys. Lett. B \textbf{385} (1996), 63-74
doi:10.1016/0370-2693(96)00874-X
[arXiv:hep-th/9606077 [hep-th]].

\bibitem{Popov:1998pc}
A.~D.~Popov,
``Selfdual Yang-Mills: Symmetries and moduli space,''
Rev. Math. Phys. \textbf{11} (1999), 1091-1149
doi:10.1142/S0129055X99000350
[arXiv:hep-th/9803183 [hep-th]].

\bibitem{Parkes:1992rz}
A.~Parkes,
``A Cubic action for selfdual Yang-Mills,''
Phys. Lett. B \textbf{286} (1992), 265-270
doi:10.1016/0370-2693(92)91773-3
[arXiv:hep-th/9203074 [hep-th]]. 

\bibitem{Cangemi:1996pf}
D.~Cangemi,
``Selfduality and maximally helicity violating QCD amplitudes,''
Int. J. Mod. Phys. A \textbf{12} (1997), 1215-1226
doi:10.1142/S0217751X97000943
[arXiv:hep-th/9610021 [hep-th]].


\bibitem{Monteiro:2011pc}
R.~Monteiro and D.~O'Connell,
``The Kinematic Algebra From the Self-Dual Sector,''
JHEP \textbf{07} (2011), 007
doi:10.1007/JHEP07(2011)007
[arXiv:1105.2565 [hep-th]].


\bibitem{Chacon:2020fmr}
E.~Chac\'on, H.~Garc\'\i{}a-Compe\'an, A.~Luna, R.~Monteiro and C.~D.~White,
``New heavenly double copies,''
JHEP \textbf{03} (2021), 247
doi:10.1007/JHEP03(2021)247
[arXiv:2008.09603 [hep-th]].

\bibitem{Campiglia:2021srh}
M.~Campiglia and S.~Nagy,
``A double copy for asymptotic symmetries in the self-dual sector,''
JHEP \textbf{03} (2021), 262
doi:10.1007/JHEP03(2021)262
[arXiv:2102.01680 [hep-th]].

\bibitem{Monteiro:2022nqt}
R.~Monteiro, R.~Stark-Much\~ao and S.~Wikeley,
``Anomaly and double copy in quantum self-dual Yang-Mills and gravity,''
[arXiv:2211.12407 [hep-th]].

\bibitem{Monteiro:2013rya}
R.~Monteiro and D.~O'Connell,
``The Kinematic Algebras from the Scattering Equations,''
JHEP \textbf{03} (2014), 110
doi:10.1007/JHEP03(2014)110
[arXiv:1311.1151 [hep-th]].

\bibitem{Cheung:2016prv}
C.~Cheung and C.~H.~Shen,
``Symmetry for Flavor-Kinematics Duality from an Action,''
Phys. Rev. Lett. \textbf{118} (2017) no.12, 121601
doi:10.1103/PhysRevLett.118.121601
[arXiv:1612.00868 [hep-th]].

\bibitem{Chen:2019ywi}
G.~Chen, H.~Johansson, F.~Teng and T.~Wang,
``On the kinematic algebra for BCJ numerators beyond the MHV sector,''
JHEP \textbf{11} (2019), 055
doi:10.1007/JHEP11(2019)055
[arXiv:1906.10683 [hep-th]].

\bibitem{Elor:2020nqe}
G.~Elor, K.~Farnsworth, M.~L.~Graesser and G.~Herczeg,
``The Newman-Penrose Map and the Classical Double Copy,''
JHEP \textbf{12} (2020), 121
doi:10.1007/JHEP12(2020)121
[arXiv:2006.08630 [hep-th]].

\bibitem{Armstrong-Williams:2022apo}
K.~Armstrong-Williams, C.~D.~White and S.~Wikeley,
``Non-perturbative aspects of the self-dual double copy,''
JHEP \textbf{08} (2022), 160
doi:10.1007/JHEP08(2022)160
[arXiv:2205.02136 [hep-th]].

\bibitem{Farnsworth:2021wvs}
K.~Farnsworth, M.~L.~Graesser and G.~Herczeg,
``Twistor space origins of the Newman-Penrose map,''
SciPost Phys. \textbf{13} (2022) no.4, 099
doi:10.21468/SciPostPhys.13.4.099
[arXiv:2104.09525 [hep-th]].

\bibitem{Skvortsov:2022unu}
E.~Skvortsov and R.~Van Dongen,
``Minimal models of field theories: SDYM and SDGR,''
JHEP \textbf{08} (2022), 083
doi:10.1007/JHEP08(2022)083
[arXiv:2204.09313 [hep-th]].

\bibitem{He:2015wgf}
S.~He, R.~Monteiro and O.~Schlotterer,
``String-inspired BCJ numerators for one-loop MHV amplitudes,''
JHEP \textbf{01} (2016), 171
doi:10.1007/JHEP01(2016)171
[arXiv:1507.06288 [hep-th]].

\bibitem{Berman:2018hwd}
D.~S.~Berman, E.~Chac\'on, A.~Luna and C.~D.~White,
``The self-dual classical double copy, and the Eguchi-Hanson instanton,''
JHEP \textbf{01} (2019), 107
doi:10.1007/JHEP01(2019)107
[arXiv:1809.04063 [hep-th]].

\bibitem{Nagy:2022xxs}
S.~Nagy and J.~Peraza,
``Radiative phase space extensions at all orders in r for self-dual Yang-Mills and gravity,''
JHEP \textbf{02} (2023), 202
doi:10.1007/JHEP02(2023)202
[arXiv:2211.12991 [hep-th]].


\bibitem{Bern:2008qj}
Z.~Bern, J.~J.~M.~Carrasco and H.~Johansson,
``New Relations for Gauge-Theory Amplitudes,''
Phys. Rev. D \textbf{78} (2008), 085011
doi:10.1103/PhysRevD.78.085011
[arXiv:0805.3993 [hep-ph]].

\bibitem{Kawai:1985xq}
H.~Kawai, D.~C.~Lewellen and S.~H.~H.~Tye,
``A Relation Between Tree Amplitudes of Closed and Open Strings,''
Nucl. Phys. B \textbf{269} (1986), 1-23
doi:10.1016/0550-3213(86)90362-7

\bibitem{Bern:2010yg}
Z.~Bern, T.~Dennen, Y.~t.~Huang and M.~Kiermaier,
``Gravity as the Square of Gauge Theory,''
Phys. Rev. D \textbf{82} (2010), 065003
doi:10.1103/PhysRevD.82.065003
[arXiv:1004.0693 [hep-th]].

\bibitem{Bern:2010ue}
Z.~Bern, J.~J.~M.~Carrasco and H.~Johansson,
``Perturbative Quantum Gravity as a Double Copy of Gauge Theory,''
Phys. Rev. Lett. \textbf{105} (2010), 061602
doi:10.1103/PhysRevLett.105.061602
[arXiv:1004.0476 [hep-th]].


\bibitem{Ward:1977ta}
R.~S.~Ward,
``On Selfdual gauge fields,''
Phys. Lett. A \textbf{61} (1977), 81-82
doi:10.1016/0375-9601(77)90842-8

\bibitem{Dunajski_1998}
M Dunajski and L J Mason and N M J Woodhouse,
``From 2D integrable systems to self-dual gravity,''
Journal of Physics A: Mathematical and General \textbf{31} 28 (1998), 6019-6028
doi:10.1088/0305-4470/31/28/015

\bibitem{Park:1989vq}
Q.~H.~Park,
``Selfdual Gravity as a Large $N$ Limit of the Two-dimensional Nonlinear $\sigma$ Model,''
Phys. Lett. B \textbf{238} (1990), 287-290
doi:10.1016/0370-2693(90)91737-V

\bibitem{Husain:1993dp}
V.~Husain,
``Selfdual gravity as a two-dimensional theory and conservation laws,''
Class. Quant. Grav. \textbf{11} (1994), 927-938
doi:10.1088/0264-9381/11/4/011
[arXiv:gr-qc/9310003 [gr-qc]].


\bibitem{Monteiro:2022lwm}
R.~Monteiro,
``Celestial chiral algebras, colour-kinematics duality and integrability,''
JHEP \textbf{01} (2023), 092
doi:10.1007/JHEP01(2023)092
[arXiv:2208.11179 [hep-th]].

\bibitem{Strominger:2021lvk}
A.~Strominger,
``w(1+infinity) and the Celestial Sphere,''
[arXiv:2105.14346 [hep-th]].

\bibitem{Guevara:2021abz}
A.~Guevara, E.~Himwich, M.~Pate and A.~Strominger,
``Holographic symmetry algebras for gauge theory and gravity,''
JHEP \textbf{11} (2021), 152
doi:10.1007/JHEP11(2021)152
[arXiv:2103.03961 [hep-th]].

\bibitem{Adamo:2021lrv}
T.~Adamo, L.~Mason and A.~Sharma,
``Celestial $w_{1+\infty}$ Symmetries from Twistor Space,''
SIGMA \textbf{18} (2022), 016
doi:10.3842/SIGMA.2022.016
[arXiv:2110.06066 [hep-th]].

\bibitem{Donnay:2022sdg}
L.~Donnay, S.~Pasterski and A.~Puhm,
``Goldilocks modes and the three scattering bases,''
JHEP \textbf{06} (2022), 124
doi:10.1007/JHEP06(2022)124
[arXiv:2202.11127 [hep-th]].

\bibitem{Pasterski:2020pdk}
S.~Pasterski and A.~Puhm,
``Shifting spin on the celestial sphere,''
Phys. Rev. D \textbf{104} (2021) no.8, 086020
doi:10.1103/PhysRevD.104.086020
[arXiv:2012.15694 [hep-th]].

\bibitem{Casali:2020vuy}
E.~Casali and A.~Puhm,
``Double Copy for Celestial Amplitudes,''
Phys. Rev. Lett. \textbf{126} (2021) no.10, 101602
doi:10.1103/PhysRevLett.126.101602
[arXiv:2007.15027 [hep-th]].

\bibitem{Donnay:2020guq}
L.~Donnay, S.~Pasterski and A.~Puhm,
``Asymptotic Symmetries and Celestial CFT,''
JHEP \textbf{09} (2020), 176
doi:10.1007/JHEP09(2020)176
[arXiv:2005.08990 [hep-th]].

\bibitem{Casali:2020uvr}
E.~Casali and A.~Sharma,
``Celestial double copy from the worldsheet,''
JHEP \textbf{05} (2021), 157
doi:10.1007/JHEP05(2021)157

\bibitem{Mason:2022hly}
L.~Mason,
``Gravity from holomorphic discs and celestial $Lw_{1+\infty}$ symmetries,''
[arXiv:2212.10895 [hep-th]].

\bibitem{Monteiro:2022xwq}
R.~Monteiro,
``From Moyal deformations to chiral higher-spin theories and to celestial algebras,''
JHEP \textbf{03} (2023), 062
doi:10.1007/JHEP03(2023)062
[arXiv:2212.11266 [hep-th]].

\bibitem{Ball:2021tmb}
A.~Ball, S.~A.~Narayanan, J.~Salzer and A.~Strominger,
``Perturbatively exact $w_{1+\infty}$ asymptotic symmetry of quantum self-dual gravity,''
JHEP \textbf{01} (2022), 114
doi:10.1007/JHEP01(2022)114
[arXiv:2111.10392 [hep-th]].

\bibitem{Adamo:2019ipt}
T.~Adamo, L.~Mason and A.~Sharma,
``Celestial amplitudes and conformal soft theorems,''
Class. Quant. Grav. \textbf{36} (2019) no.20, 205018
doi:10.1088/1361-6382/ab42ce
[arXiv:1905.09224 [hep-th]].

\bibitem{Strominger:2013jfa}
A.~Strominger,
``On BMS Invariance of Gravitational Scattering,''
JHEP \textbf{07} (2014), 152
doi:10.1007/JHEP07(2014)152
[arXiv:1312.2229 [hep-th]].

\bibitem{He:2014laa}
T.~He, V.~Lysov, P.~Mitra and A.~Strominger,
``BMS supertranslations and Weinberg\textquoteright{}s soft graviton theorem,''
JHEP \textbf{05} (2015), 151
doi:10.1007/JHEP05(2015)151
[arXiv:1401.7026 [hep-th]].


\bibitem{McLoughlin:2022ljp}
T.~McLoughlin, A.~Puhm and A.~M.~Raclariu,
``The SAGEX review on scattering amplitudes chapter 11: soft theorems and celestial amplitudes,''
J. Phys. A \textbf{55} (2022) no.44, 443012
doi:10.1088/1751-8121/ac9a40
[arXiv:2203.13022 [hep-th]].

\bibitem{Kampf:2023elx}
K.~Kampf, J.~Novotny, J.~Trnka and P.~Vasko,
``Goldstone bosons on celestial sphere and conformal soft theorems,''
[arXiv:2303.14761 [hep-th]].

\bibitem{Freidel:2021ytz}
L.~Freidel, D.~Pranzetti and A.~M.~Raclariu,
``Higher spin dynamics in gravity and w1+\ensuremath{\infty} celestial symmetries,''
Phys. Rev. D \textbf{106} (2022) no.8, 086013
doi:10.1103/PhysRevD.106.086013
[arXiv:2112.15573 [hep-th]].


\bibitem{Maldacena:1997re}
J.~M.~Maldacena,
``The Large N limit of superconformal field theories and supergravity,''
Adv. Theor. Math. Phys. \textbf{2}, 231-252 (1998)
doi:10.1023/A:1026654312961
[arXiv:hep-th/9711200 [hep-th]].

\bibitem{Strominger:2001gp}
A.~Strominger,
``Inflation and the dS / CFT correspondence,''
JHEP \textbf{11} (2001), 049
doi:10.1088/1126-6708/2001/11/049
[arXiv:hep-th/0110087 [hep-th]].

\bibitem{Maldacena:2002vr}
J.~M.~Maldacena,
``Non-Gaussian features of primordial fluctuations in single field inflationary models,''
JHEP \textbf{05} (2003), 013
doi:10.1088/1126-6708/2003/05/013
[arXiv:astro-ph/0210603 [astro-ph]].

\bibitem{McFadden:2009fg}
P.~McFadden and K.~Skenderis,
``Holography for Cosmology,''
Phys. Rev. D \textbf{81} (2010), 021301
doi:10.1103/PhysRevD.81.021301
[arXiv:0907.5542 [hep-th]].


\bibitem{4ptdouble}
C.~Armstrong, H.~Goodhew, A.~Lipstein and J.~Mei,
``Graviton Trispectrum from Gluons,''
[arXiv:2304.07206 [hep-th]].

\bibitem{Armstrong:2020woi}
C.~Armstrong, A.~E.~Lipstein and J.~Mei,
``Color/kinematics duality in AdS$_{4}$,''
JHEP \textbf{02} (2021), 194
doi:10.1007/JHEP02(2021)194
[arXiv:2012.02059 [hep-th]].

\bibitem{Albayrak:2020fyp}
S.~Albayrak, S.~Kharel and D.~Meltzer,
``On duality of color and kinematics in (A)dS momentum space,''
JHEP \textbf{03} (2021), 249
doi:10.1007/JHEP03(2021)249
[arXiv:2012.10460 [hep-th]].

\bibitem{Alday:2021odx}
L.~F.~Alday, C.~Behan, P.~Ferrero and X.~Zhou,
``Gluon Scattering in AdS from CFT,''
JHEP \textbf{06} (2021), 020
doi:10.1007/JHEP06(2021)020
[arXiv:2103.15830 [hep-th]].

\bibitem{Diwakar:2021juk}
P.~Diwakar, A.~Herderschee, R.~Roiban and F.~Teng,
``BCJ amplitude relations for Anti-de Sitter boundary correlators in embedding space,''
JHEP \textbf{10} (2021), 141
doi:10.1007/JHEP10(2021)141
[arXiv:2106.10822 [hep-th]].

\bibitem{Sivaramakrishnan:2021srm}
A.~Sivaramakrishnan,
``Towards color-kinematics duality in generic spacetimes,''
JHEP \textbf{04} (2022), 036
doi:10.1007/JHEP04(2022)036
[arXiv:2110.15356 [hep-th]].

\bibitem{Cheung:2022pdk}
C.~Cheung, J.~Parra-Martinez and A.~Sivaramakrishnan,
``On-shell correlators and color-kinematics duality in curved symmetric spacetimes,''
JHEP \textbf{05} (2022), 027
doi:10.1007/JHEP05(2022)027
[arXiv:2201.05147 [hep-th]].

\bibitem{Herderschee:2022ntr}
A.~Herderschee, R.~Roiban and F.~Teng,
``On the differential representation and color-kinematics duality of AdS boundary correlators,''
JHEP \textbf{05} (2022), 026
doi:10.1007/JHEP05(2022)026
[arXiv:2201.05067 [hep-th]].

\bibitem{Drummond:2022dxd}
J.~M.~Drummond, R.~Glew and M.~Santagata,
``Bern-Carrasco-Johansson relations in AdS5\texttimes{}S3 and the double-trace spectrum of super gluons,''
Phys. Rev. D \textbf{107} (2023) no.8, L081901
doi:10.1103/PhysRevD.107.L081901
[arXiv:2202.09837 [hep-th]].

\bibitem{Farrow:2018yni}
J.~A.~Farrow, A.~E.~Lipstein and P.~McFadden,
``Double copy structure of CFT correlators,''
JHEP \textbf{02} (2019), 130
doi:10.1007/JHEP02(2019)130
[arXiv:1812.11129 [hep-th]].

\bibitem{Lipstein:2019mpu}
A.~E.~Lipstein and P.~McFadden,
``Double copy structure and the flat space limit of conformal correlators in even dimensions,''
Phys. Rev. D \textbf{101} (2020) no.12, 125006
doi:10.1103/PhysRevD.101.125006
[arXiv:1912.10046 [hep-th]].

\bibitem{Jain:2021qcl}
S.~Jain, R.~R.~John, A.~Mehta, A.~A.~Nizami and A.~Suresh,
``Double copy structure of parity-violating CFT correlators,''
JHEP \textbf{07} (2021), 033
doi:10.1007/JHEP07(2021)033
[arXiv:2104.12803 [hep-th]].

\bibitem{Zhou:2021gnu}
X.~Zhou,
``Double Copy Relation in AdS Space,''
Phys. Rev. Lett. \textbf{127} (2021) no.14, 141601
doi:10.1103/PhysRevLett.127.141601
[arXiv:2106.07651 [hep-th]].

\bibitem{Armstrong:2022csc}
C.~Armstrong, H.~Gomez, R.~Lipinski Jusinskas, A.~Lipstein and J.~Mei,
``Effective field theories and cosmological scattering equations,''
JHEP \textbf{08} (2022), 054
doi:10.1007/JHEP08(2022)054
[arXiv:2204.08931 [hep-th]].

\bibitem{Adamo:2017nia}
T.~Adamo, E.~Casali, L.~Mason and S.~Nekovar,
``Scattering on plane waves and the double copy,''
Class. Quant. Grav. \textbf{35} (2018) no.1, 015004
doi:10.1088/1361-6382/aa9961
[arXiv:1706.08925 [hep-th]].

\bibitem{Adamo:2018mpq}
T.~Adamo, E.~Casali, L.~Mason and S.~Nekovar,
``Plane wave backgrounds and colour-kinematics duality,''
JHEP \textbf{02} (2019), 198
doi:10.1007/JHEP02(2019)198
[arXiv:1810.05115 [hep-th]]. 

\bibitem{Bahjat-Abbas:2017htu}
N.~Bahjat-Abbas, A.~Luna and C.~D.~White,
``The Kerr-Schild double copy in curved spacetime,''
JHEP \textbf{12} (2017), 004
doi:10.1007/JHEP12(2017)004
[arXiv:1710.01953 [hep-th]].

\bibitem{Carrillo-Gonzalez:2017iyj}
M.~Carrillo-Gonz\'alez, R.~Penco and M.~Trodden,
``The classical double copy in maximally symmetric spacetimes,''
JHEP \textbf{04} (2018), 028
doi:10.1007/JHEP04(2018)028
[arXiv:1711.01296 [hep-th]].

\bibitem{Borsten:2019prq}
L.~Borsten, I.~Jubb, V.~Makwana and S.~Nagy,
``Gauge \texttimes{} gauge on spheres,''
JHEP \textbf{06} (2020), 096
doi:10.1007/JHEP06(2020)096
[arXiv:1911.12324 [hep-th]].

\bibitem{Borsten:2021zir}
L.~Borsten, I.~Jubb, V.~Makwana and S.~Nagy,
``Gauge \texttimes{} gauge = gravity on homogeneous spaces using tensor convolutions,''
JHEP \textbf{06} (2021), 117
doi:10.1007/JHEP06(2021)117
[arXiv:2104.01135 [hep-th]].

\bibitem{Chawla:2022ogv}
S.~Chawla and C.~Keeler,
``Aligned fields double copy to Kerr-NUT-(A)dS,''
JHEP \textbf{04} (2023), 005
doi:10.1007/JHEP04(2023)005
[arXiv:2209.09275 [hep-th]].

\bibitem{explEq6}
We note that it is possible to give a more covariant looking expression for the components in \eqref{YM flat in comp}, see  Section 2 of \cite{Campiglia:2021srh}.

\bibitem{Sonego:1993fw}
S.~Sonego and V.~Faraoni,
``Coupling to the curvature for a scalar field from the equivalence principle,''
Class. Quant. Grav. \textbf{10}, 1185-1187 (1993)
doi:10.1088/0264-9381/10/6/015



\bibitem{Raju:2011mp}
S.~Raju,
``Recursion Relations for AdS/CFT Correlators,''
Phys. Rev. D \textbf{83} (2011), 126002
doi:10.1103/PhysRevD.83.126002
[arXiv:1102.4724 [hep-th]].

\bibitem{Maldacena:2011nz}
J.~M.~Maldacena and G.~L.~Pimentel,
``On graviton non-Gaussianities during inflation,''
JHEP \textbf{09} (2011), 045
doi:10.1007/JHEP09(2011)045
[arXiv:1104.2846 [hep-th]].

\bibitem{Bzowski:2013sza}
A.~Bzowski, P.~McFadden and K.~Skenderis,
``Implications of conformal invariance in momentum space,''
JHEP \textbf{03} (2014), 111
doi:10.1007/JHEP03(2014)111
[arXiv:1304.7760 [hep-th]].

\bibitem{Bzowski:2015pba}
A.~Bzowski, P.~McFadden and K.~Skenderis,
``Scalar 3-point functions in CFT: renormalisation, beta functions and anomalies,''
JHEP \textbf{03} (2016), 066
doi:10.1007/JHEP03(2016)066
[arXiv:1510.08442 [hep-th]].


\bibitem{Fairlie:1990wv}
D.~B.~Fairlie and J.~Nuyts,
``Deformations and Renormalizations of $W$(infinity),''
Commun. Math. Phys. \textbf{134} (1990), 413-420
doi:10.1007/BF02097709

\bibitem{Ward:1980am}
R.~S.~Ward,
Commun. Math. Phys. \textbf{78} (1980), 1-17
doi:10.1007/BF01941967


\bibitem{Krasnov:2016emc}
K.~Krasnov,
``Self-Dual Gravity,''
Class. Quant. Grav. \textbf{34} (2017) no.9, 095001
doi:10.1088/1361-6382/aa65e5
[arXiv:1610.01457 [hep-th]].

\bibitem{Krasnov:2021cva}
K.~Krasnov and E.~Skvortsov,
``Flat self-dual gravity,''
JHEP \textbf{08} (2021), 082
doi:10.1007/JHEP08(2021)082
[arXiv:2106.01397 [hep-th]].

\bibitem{Neiman:2023bkq}
Y.~Neiman,
``Self-dual gravity in de Sitter space: lightcone ansatz and static-patch scattering,''
[arXiv:2303.17866 [gr-qc]].

\bibitem{Krasnov:2017dww}
K.~Krasnov,
``Field redefinitions and Plebanski formalism for GR,''
Class. Quant. Grav. \textbf{35} (2018) no.14, 147001
doi:10.1088/1361-6382/aac844
[arXiv:1708.07694 [gr-qc]].

\bibitem{Krasnov:2021nsq}
K.~Krasnov, E.~Skvortsov and T.~Tran,
``Actions for self-dual Higher Spin Gravities,''
JHEP \textbf{08} (2021), 076
doi:10.1007/JHEP08(2021)076
[arXiv:2105.12782 [hep-th]].

\bibitem{Krasnov:2022mvn}
K.~Krasnov and A.~Shaw,
``Weyl curvature evolution system for GR,''
Class. Quant. Grav. \textbf{40} (2023) no.7, 075013
doi:10.1088/1361-6382/acc0cc
[arXiv:2212.12273 [gr-qc]].

\bibitem{Herfray:2022prf}
Y.~Herfray, K.~Krasnov and E.~Skvortsov,
``Higher-spin self-dual Yang-Mills and gravity from the twistor space,''
JHEP \textbf{01} (2023), 158
doi:10.1007/JHEP01(2023)158
[arXiv:2210.06209 [hep-th]].

\bibitem{Przanowski:1983xpa}
M.~Przanowski,
``LOCALLY HERMITE EINSTEIN, SELFDUAL GRAVITATIONAL INSTANTONS,''
Acta Phys. Polon. B \textbf{14} (1983), 625-627


\bibitem{Adamo:2021bej}
T.~Adamo, L.~Mason and A.~Sharma,
``Twistor sigma models for quaternionic geometry and graviton scattering,''
[arXiv:2103.16984 [hep-th]].


\bibitem{Donnay:2022aba}
L.~Donnay, A.~Fiorucci, Y.~Herfray and R.~Ruzziconi,
``Carrollian Perspective on Celestial Holography,''
Phys. Rev. Lett. \textbf{129} (2022) no.7, 071602
doi:10.1103/PhysRevLett.129.071602
[arXiv:2202.04702 [hep-th]].

\bibitem{deGioia:2023cbd}
L.~P.~de Gioia and A.~M.~Raclariu,
``Celestial Sector in CFT: Conformally Soft Symmetries,''
[arXiv:2303.10037 [hep-th]].

\bibitem{Banerjee:2022oll}
N.~Banerjee, K.~Fernandes and A.~Mitra,
``$1/L^2$ corrected soft photon theorem from a CFT$_3$ Ward identity,''
[arXiv:2209.06802 [hep-th]].

\bibitem{Bagchi:2023fbj}
A.~Bagchi, P.~Dhivakar and S.~Dutta,
``AdS Witten Diagrams to Carrollian Correlators,''
[arXiv:2303.07388 [hep-th]].

\bibitem{Saha:2023hsl}
A.~Saha,
``Carrollian Approach to $1+3$D Flat Holography,''
[arXiv:2304.02696 [hep-th]].


\bibitem{Bern:1998xc}
Z.~Bern, L.~J.~Dixon, M.~Perelstein and J.~S.~Rozowsky,
``One loop n point helicity amplitudes in (selfdual) gravity,''
Phys. Lett. B \textbf{444} (1998), 273-283
doi:10.1016/S0370-2693(98)01397-5
[arXiv:hep-th/9809160 [hep-th]].

\bibitem{Bern:1993qk}
Z.~Bern, G.~Chalmers, L.~J.~Dixon and D.~A.~Kosower,
``One loop N gluon amplitudes with maximal helicity violation via collinear limits,''
Phys. Rev. Lett. \textbf{72} (1994), 2134-2137
doi:10.1103/PhysRevLett.72.2134
[arXiv:hep-ph/9312333 [hep-ph]].

\bibitem{Boels:2013bi}
R.~H.~Boels, R.~S.~Isermann, R.~Monteiro and D.~O'Connell,
``Colour-Kinematics Duality for One-Loop Rational Amplitudes,''
JHEP \textbf{04} (2013), 107
doi:10.1007/JHEP04(2013)107
[arXiv:1301.4165 [hep-th]].


\bibitem{Sharapov:2022awp}
A.~Sharapov and E.~Skvortsov,
``Chiral higher spin gravity in (A)dS4 and secrets of Chern\textendash{}Simons matter theories,''
Nucl. Phys. B \textbf{985} (2022), 115982
doi:10.1016/j.nuclphysb.2022.115982
[arXiv:2205.15293 [hep-th]].

\bibitem{aknMP}
We thank Malcolm Perry for this observation.



\bibitem{Pope:1991ig}
C.~N.~Pope,
``Lectures on W algebras and W gravity,''
[arXiv:hep-th/9112076 [hep-th]].

\bibitem{Bittleston:2023bzp}
R.~Bittleston, S.~Heuveline and D.~Skinner,
``The Celestial Chiral Algebra of Self-Dual Gravity on Eguchi-Hanson Space,''
[arXiv:2305.09451 [hep-th]].

\bibitem{Bu:2022iak}
W.~Bu, S.~Heuveline and D.~Skinner,
JHEP \textbf{12} (2022), 011
doi:10.1007/JHEP12(2022)011
[arXiv:2208.13750 [hep-th]].

\bibitem{Etnigof:2020xx}
P.~Etingof, D.~Kalinov and E.~Rains
``New realizations of deformed double current algebras and Deligne categories"
[arXiv:2005.13604 [math.RT]].





\end{thebibliography}
\end{document}